\documentclass[a4paper,11pt]{article}
\pdfoutput=1 

\usepackage{jheppub} 
\usepackage{amsmath}
\usepackage{pict2e}
\usepackage{physics}
\usepackage{amsfonts}
\usepackage{mathtools}
\usepackage{amssymb,bbm}
\usepackage{dsfont}
\usepackage{subcaption}
\usepackage{braket}
\usepackage{amsthm}
\usepackage{xcolor}
\usepackage{tikz}
\usepackage{ytableau}
\usepackage{graphicx}
\usetikzlibrary{arrows.meta}
\usepackage{hyperref}

\makeatletter
\newcommand{\farsquare}[2]{#1\,{\mathpalette\far@square{#2}}}
\newcommand{\far@square}[2]{%
  \mathop{\vcenter{\hbox{%
    \sbox\z@{$\m@th#1\sum$}%
    \setlength{\unitlength}{0.9\dimexpr\ht\z@+\dp\z@}%
    \begin{picture}(1,1)
    \roundjoin
    \polyline(0,0)(0,1)(1,1)(1,0)(0,0)(0,0.5)
    \end{picture}%
  }}}\limits_{#1#2}%
}
\makeatother

\newcommand{\nc}{\newcommand}
\nc{\rnc}{\renewcommand} 

\rnc{\a}{\alpha}
\rnc{\b}{\beta}
\nc{\g}{\gamma}
\rnc{\d}{\delta}
\nc{\e}{\epsilon}
\nc{\ee}{\varepsilon}
\nc{\z}{\zeta}
\nc{\f}{\phi}
\nc{\m}{\mu}
\nc{\n}{\nu}
\rnc{\r}{\rho}
\rnc{\k}{\kappa}
\rnc{\l}{\lambda}
\nc{\p}{\pi}
\nc{\s}{\sigma}
\rnc{\t}{\tau}
\nc{\w}{\omega}
\nc{\x}{\chi}
\nc{\F}{\Phi}
\rnc{\L}{\Lambda}

\nc{\1}{\mathbbm{1}}

\newcommand{\cM}{\mathcal{M}}

\newcommand{\cI}{\mathcal{I}}

\def\CI{{\cal I}}

\usetikzlibrary{decorations.markings}

\usepackage[draft,inline,nomargin,index]{fixme}
\FXRegisterAuthor{kr}{akr}{\color{blue}KR}

\title{\boldmath Defect relative entropy in symmetric  orbifold CFTs}

\author[a]{Mostafa Ghasemi}

\affiliation[a]{
School of physics, Institute for Research in Fundamental Sciences (IPM) \\
P.O.Box 19395-5531, Tehran, Iran.
}

\emailAdd{ghasemi.mg@ipm.ir}

\abstract{
In this work, we compute the defect relative entropy between topological defects in the symmetric product orbifold CFT $\mathrm{Sym}^N(M) = M^{\otimes N}/S_N$. Our analysis covers two distinct classes of defects: universal defects, which realize the $\mathrm{Rep}(S_N)$ non-invertible symmetry, and non-universal defects.

We show that the defect relative entropy reduces to a Kullback--Leibler (KL) divergence. The resulting expression decomposes naturally into two contributions: one governed by characters of the symmetric group $S_N$, and the other controlled by modular $S$-matrix elements of the seed RCFT. Remarkably, both sets of data appear as probability distributions, yielding an information-theoretic interpretation of permutation group data and modular data within the symmetric orbifold.

The structure of the divergence depends sensitively on the defect class. For universal defects, only the permutation group data contributes; for maximally fractional defects, both permutation and modular data enter and together define the relevant probability distributions. This feature suggests that the maximally fractional defect can be understood as a kind of product of the RCFT defect and the symmetric orbifold defect.
 }

\begin{document} 
\maketitle
\flushbottom

\section{Introduction}
\label{sec:intro}

The orbifold construction in two-dimensional conformal field theory (CFT) provides a powerful method for deriving new CFTs from old ones \cite{Dixon:1985jw, Dixon:1986qv, Dijkgraaf:1989hb}. Among these, symmetric product orbifolds \cite{Klemm:1990df, Bantay:2000eq, Lunin:2000yv, Dijkgraaf:1996xw, Pakman:2009zz} are particularly remarkable: they constitute a rare class of large-$N$ theories that remain exactly solvable while capturing essential features of strongly coupled quantum gravity through the AdS/CFT correspondence. The most well-studied example is the duality between the tensionless string theory on $\rm AdS_3\times S^3\times\mathbb{T}^4$ and the symmetric orbifold of $\mathbb{T}^4$, denoted by ${\rm Sym}^N(\mathbb{T}^4)$ \cite{Gaberdiel:2018rqv, Eberhardt:2018ouy, Giribet:2018ada, Eberhardt:2019ywk}. 
The theory $\mathrm{Sym}^N(\mathcal{M})$ is constructed by taking $N$ copies of a seed CFT $\mathcal{M}$ and gauging the permutation symmetry $S_N$:
\begin{equation}
	\mathrm{Sym}^N(\mathcal{M}) := \mathcal{M}^{\otimes N} / S_N ,
\end{equation}
resulting in a spectrum organized into twisted sectors labeled by conjugacy classes of.

	Beyond their holographic significance, symmetric product orbifolds provide a powerful setting for studying \emph{topological defects} and the modern framework of \emph{non-invertible symmetries}. In two dimensions, such symmetries are implemented by topological defect lines (TDLs)—codimension-one operators that commute with the Virasoro algebra and can be freely deformed in correlation functions, provided they do not cross local operator insertions~\cite{Verlinde:1988sn, Petkova:2000ip, Petkova:2001ag, Fuchs:2002cm, Frohlich:2006ch, Frohlich:2009gb}. 
	
	While invertible defects correspond to conventional global symmetries, generic defects are non-invertible and obey a fusion product that generalizes group multiplication. This yields a unified description of generalized global symmetries in quantum field theory \cite{Gaiotto:2014kfa}.\footnote{For comprehensive reviews, see~\cite{Shao:2023gho, Bhardwaj:2023kri, Schafer-Nameki:2023jdn}.}
	The study of these defects in orbifold theories has provided deep insights into renormalization group flows, symmetry topological field theories, and generalized gauging/orbifolding procedures~\cite{Gaiotto:2012np, Bhardwaj:2017xup, Tachikawa:2017gyf, Chang:2018iay, Thorngren:2019iar, Gaiotto:2020iye, Thorngren:2021yso,Kaidi:2021xfk, Choi:2023vgk, Perez-Lona:2023djo, Diatlyk:2023fwf, Perez-Lona:2024sds, Lu:2025gpt}. These developments establish a foundational link between algebraic structures and non-perturbative dynamics.

Symmetric product orbifolds naturally exhibit a rich spectrum of non-invertible symmetries, inherited from the gauging of the permutation group $S_N$. Defects associated with individual conjugacy classes—as well as formal linear combinations thereof—are generically non-invertible~\cite{Gutperle:2024vyp, Knighton:2024noc, Benjamin:2025knd}.

A particularly insightful probe of non-invertible symmetries is provided by quantum entanglement~\cite{Choi:2024wfm}. The entanglement entropy of a subregion is affected not only by local degrees of freedom but also by global and topological data. This is demonstrated when a topological defect intersects an entangling surface, which is introducing defect-specific contributions to the entropy~\cite{Sakai:2008tt,Brehm:2015plf, Gutperle:2015kmw}. In two-dimensional conformal field theories (CFTs)~\cite{Sakai:2008tt,Brehm:2015plf, Gutperle:2015kmw, Brehm:2015lja, Nishioka:2021cxe, Northe:2025zmv, Northe:2025qcv} and in the context of holography \cite{ Gutperle:2015hcv, Karch:2021qhd, Karch:2024udk}, these contributions are systematically computed via the replica trick, defining a concept of defect entanglement entropy. For symmetric product orbifolds, this entropy can be interpreted  in terms of permutation statistics~\cite{ Gutperle:2024rvo}.

While entanglement entropy has proven to be a powerful tool across quantum many-body physics, quantum field theory, and holography~\cite{Calabrese:2009qy, Nishioka:2018khk, Casini:2022rlv}, it suffers from ultraviolet divergences and subtle definitional ambiguities in gauge theories. A more refined information-theoretic measure, free from these ambiguities, is \emph{quantum relative entropy}~\cite{Casini:2013rba}.

Relative entropy, \(D(\rho\|\sigma)\), quantifies the statistical distinguishability between two quantum states \(\rho\) and \(\sigma\) and is defined as
\begin{equation}
	D(\rho\|\sigma)
	= \operatorname{tr}(\rho \log \rho)
	- \operatorname{tr}(\rho \log \sigma),
	\label{eq:relative_entropy}
\end{equation}
satisfying positivity \(D(\rho\|\sigma)\geq 0\) with equality if and only if \(\rho=\sigma\). Its utility extends across quantum information theory \cite{Vedral:2002zz}, statistical mechanics \cite{Wehrl:1978zz}, and quantum field theory. Key properties like positivity and monotonicity under completely positive trace-preserving maps have established it as a powerful probe in diverse contexts, including quantum field theory and RG flow \cite{Casini:2016fgb, Casini:2016udt, Casini:2018nym, Casini:2019qst, Casini:2020rgj, Casini:2022bsu}, conformal field theory (CFT) \cite{Lashkari:2014yva,Lashkari:2015dia, Ruggiero:2016khg}, boundary conformal field theory \cite{ Ghasemi:2024wcq}, holography \cite{Blanco:2013joa, Wong:2013gua,Jafferis:2015del}, quantum gravity \cite{Casini:2008cr,Wall:2011hj, Faulkner:2016mzt, Balakrishnan:2017bjg,Longo:2018zib}, and study of random states \cite{Kudler-Flam:2021rpr, Ghasemi:2024yzw} . A useful one-parameter generalization is the sandwiched R\'enyi relative entropy \cite{Tomamichel:2013bde,Wilde:2013bdg},
\begin{equation}
	D_n(\rho\|\sigma)
	= \frac{1}{n-1}
	\log\,
	\operatorname{tr}
	\left[
	\left(
	\sigma^{\frac{1-n}{2n}}
	\rho
	\sigma^{\frac{1-n}{2n}}
	\right)^n
	\right],
	\label{eq:renyi_relative_entropy}
\end{equation}
whose \(n\to1\) limit recovers Eq.~\eqref{eq:relative_entropy}. In practice, relative entropy between reduced density matrices in a spatial subregion can be computed via a replica trick \cite{Lashkari:2015dia},
\begin{equation}
	D(\rho\|\sigma)
	=
	-\partial_n
	\log
	\left(
	\frac{\operatorname{tr}(\rho\,\sigma^{n-1})}{\operatorname{tr}(\rho^n)}
	\right)
	\Big|_{n\to1}	=
	-\partial_n
	\log
	\left(
	\frac{Z_n(\rho\,,\sigma)}{Z_n(\rho)}
	\right)
	\Big|_{n\to1}.
	\label{eq:replica}
\end{equation}

In the context of two-dimensional CFTs, this replica method has been applied systematically to compute relative entropy between reduced density matrices associated with arbitrary excited states in a spatial subregion \cite{Lashkari:2014yva, Lashkari:2015dia}. The present work extends these techniques to the study of states and defects in symmetric product orbifold CFTs.

Recently, the concept of defect relative entropy was introduced in the context of topological interfaces in 2D CFT \cite{Ghasemi:2026hoa}, providing a well-defined measure of distinguishability in the space of topological defects themselves. This quantity sharpens the interplay between non-invertible symmetries and quantum information in CFTs. The primary goal of this paper is to compute the defect relative entropy for topological defects in symmetric product orbifold CFTs. Building on methods developed for rational CFTs \cite{Sakai:2008tt, Brehm:2015plf, Gutperle:2015kmw} and recent advances in orbifold defect technology \cite{Gutperle:2024vyp, Knighton:2024noc, Benjamin:2025knd}, we analyze both universal and non-universal families of defects. Our results illuminate the information-theoretic structure of non-invertible symmetries in a solvable, holographically relevant setting, and expose universal features independent of the choice of seed theory.

%The main goal of this paper is to calculate the defect relative entropy in the space of topological defect operators in the symmetric productorbifold CFTs \cite{Gutperle:2024vyp,Knighton:2024noc, Benjamin:2025knd}', using the methods developed for topological defects in rational CFTs in \cite{Sakai:2008tt,Brehm:2015plf,Gutperle:2015kmw}.' The defect relative entropy provide a measure of the distinguishability in the space of defects.

The paper is organized as follows. In Section~\ref{sec:topological-defects} we briefly review the construction of topological defects in 2d CFT and in symmetric product orbifolds. In Section~\ref{sec:orbifold-defect relative}, we first summarize the computation of defect relative entropy for topological defects in 2d CFT, and then calculate it for both universal and non-universal defect classes  in symmetric product orbifolds. We conclude with a discussion of implications and future directions in Section~\ref{sec:Discussion}.

\section{Topological Defects}
\label{sec:topological-defects}

%In this section, we briefly review the construction of topological defects in conformal field theory and symmetric product orbifolds. For more details in this context, see \cite{ReferenceToBeAdded}.

%This section provides a concise review of topological defects in two-dimensional conformal field theories (CFTs), with a focus on their realization in symmetric product orbifolds. For a more comprehensive treatment, see \cite{ReviewReference}.

This section provides a concise review of topological defects in two-dimensional conformal field theories (CFTs), with a focus on their realization in symmetric product orbifolds. In two-dimensional CFTs, topological defects are one-dimensional extended operators that commute with the holomorphic and antiholomorphic components of the stress-energy tensor, making them invariant under continuous deformations. In rational CFTs (RCFTs), they are classified by bimodule categories over the chiral fusion category of the theory \cite{Frohlich:2006ch}. A particularly rich setting for their study is the symmetric product orbifold $\text{Sym}^N(X) = X^N / S_N$, where the permutation symmetry gives rise to a wide class of defects that implement twists and projections among the orbifold sectors. These defects play a central role in organizing the orbifold spectrum, computing entanglement entropy, and understanding holographic dualities. For a comprehensive review of the general theory of topological defects in CFT, see \cite{Verlinde:1988sn, Petkova:2000ip, Petkova:2001ag, Fuchs:2002cm, Frohlich:2006ch, Frohlich:2009gb, Chang:2018iay}. For their specific construction and role in symmetric product orbifolds, we refer to \cite{Gutperle:2024vyp, Knighton:2024noc, Benjamin:2025knd}.

\subsection{Topological Defects in 2d CFT}
\label{subsec:2d-cft-defects}

Consider two CFTs, denoted CFT\(^{(1)}\) and CFT\(^{(2)}\), defined on regions that meet along a line $\mathcal{C}$. To preserve part of the conformal symmetry, the energy flow across $\mathcal{C}$ must be continuous. This is enforced by the \emph{conformal gluing condition}:
\begin{equation}
	\left. \left( T^{(1)} - \bar{T}^{(1)} \right) \right|_{\mathcal{C}} = \left. \left( T^{(2)} - \bar{T}^{(2)} \right) \right|_{\mathcal{C}},
	\label{eq:energy_continuity}
\end{equation}
where $T^{(k)}$ and $\bar{T}^{(k)}$ are the holomorphic and anti-holomorphic components of the stress-energy tensor in the $k$-th theory.

The line $\mathcal{C}$ can be viewed as a defect operator $\mathcal{I}$ that intertwines the Hilbert space $\mathcal{H}^{(1)}$ of the first theory with $\mathcal{H}^{(2)}$ of the second. Condition \eqref{eq:energy_continuity} implies that this operator satisfies the commutation relations:
\begin{equation}
	\left( L_n^{(1)} - \bar{L}_n^{(1)} \right) \mathcal{I} = \mathcal{I} \left( L_n^{(2)} - \bar{L}_n^{(2)} \right),
	\label{eq:weak_condition}
\end{equation}
where $L_n^{(k)}, \bar{L}_n^{(k)}$ are the Virasoro generators.

The general solution to \eqref{eq:weak_condition} is difficult to characterize. The problem simplifies considerably if the defect satisfies the stronger conditions:
\begin{equation}
	L_n^{(1)} \mathcal{I} = \mathcal{I} L_n^{(2)}, \qquad \bar{L}_n^{(1)} \mathcal{I} = \mathcal{I} \bar{L}_n^{(2)}.
	\label{eq:strong_condition}
\end{equation}
Since $\mathcal{I}$ then commutes with all Virasoro generators, its location can be moved freely on the plane without affecting correlation functions. Defects satisfying \eqref{eq:strong_condition} are therefore \emph{topological}.

Topological defects admit a natural fusion operation. Given two parallel defect lines $\mathcal{I}_a$ and $\mathcal{I}_b$, their operator product defines the fusion product
\begin{equation}
	\mathcal{I}_a \times \mathcal{I}_b \simeq \bigoplus_{c} N_{ab}^{c} \, \mathcal{I}_c,
	\label{eq:fusion-rule}
\end{equation}
where $N_{ab}^{c} \in \mathbb{Z}_{\geq 0}$ are fusion coefficients. The set of simple TDLs equipped with this fusion product forms a \emph{fusion category} $\mathcal{F}$ that encodes the symmetry structure of the CFT.

Ordinary finite group symmetries $G$ correspond to the special case where every simple defect $\mathcal{I}_g$ ($g \in G$) is invertible:
\begin{equation}
	\mathcal{I}_{g_1} \times \mathcal{I}_{g_2} \simeq \mathcal{I}_{g_1 g_2}, \qquad 
	\mathcal{I}_g \times \mathcal{I}_{g^{-1}} \simeq \mathbf{1},
\end{equation}
where $\mathbf{1}$ denotes the trivial (identity) defect. In this case, $\mathcal{F}$ is equivalent to the category $\mathsf{Rep}(G)$ of finite-dimensional representations of $G$.

\subsubsection{Topological Interfaces in Rational CFT}
Topological defects are particularly well-understood in rational CFTs (RCFTs). The Hilbert space of an RCFT decomposes into a finite sum of products of irreducible Virasoro representations:
\begin{equation}
	\mathcal{H} = \bigoplus_{i,\bar{j}} \mathcal{M}_{i\bar{j}} \, \mathcal{V}_i \otimes \bar{\mathcal{V}}_{\bar{j}},
	\label{eq:RCFT_decomposition}
\end{equation}
where $i, \bar{j}$ label the irreducible representations $\mathcal{V}_i, \bar{\mathcal{V}}_{\bar{j}}$, and $\mathcal{M}_{i\bar{j}}$ is the multiplicity of the pair $(i, \bar{j})$.

For two RCFTs with multiplicity matrices $\mathcal{M}^{(1)}_{i\bar{j}}$ and $\mathcal{M}^{(2)}_{i\bar{j}}$, topological interface operators between them can be constructed explicitly \cite{Petkova:2000ip,Petkova:2001ag,Brehm:2015plf}:
\begin{equation}
	\mathcal{I}_K = \sum_{\mathbf{i}} d_{K\mathbf{i}} \, P^{\mathbf{i}}.
	\label{eq:interface_operator}
\end{equation}
Here $K$ labels the type of topological interface. The index $\mathbf{i} \equiv (i, \bar{j}; \alpha, \beta)$ runs over representations and multiplicity labels, with $\alpha = 1,\dots,\mathcal{M}^{(1)}_{i\bar{j}}$ and $\beta = 1,\dots,\mathcal{M}^{(2)}_{i\bar{j}}$. The operator $P^{\mathbf{i}}$ is a projector intertwining the representation $(\mathcal{V}_i \otimes \bar{\mathcal{V}}_{\bar{j}})^{(\alpha)}$ of the first theory with $(\mathcal{V}_i \otimes \bar{\mathcal{V}}_{\bar{j}})^{(\beta)}$ of the second.

When the two CFTs are isomorphic, i.e., $\mathcal{M}^{(1)}_{i\bar{j}} = \mathcal{M}^{(2)}_{i\bar{j}}$, we are describing a conformal interface within the same theory. In this case, the projector can be realized as:
\begin{equation}
	P^{(i,\bar{j}; \alpha, \beta)} \equiv \sum_{n, \bar{n}} \left( |i, n\rangle \otimes |\bar{j}, \bar{n}\rangle \right)^{(\alpha)} \left( \langle i, n| \otimes \langle \bar{j}, \bar{n}| \right)^{(\beta)},
	\label{eq:projector_realization}
\end{equation}
where $\{ |i, n\rangle \otimes |\bar{j}, \bar{n}\rangle \}$ is an orthonormal basis for $\mathcal{V}_i \otimes \bar{\mathcal{V}}_{\bar{j}}$. The operator $\mathcal{I}_K$ defined in \eqref{eq:interface_operator} satisfies the strong topological conditions \eqref{eq:strong_condition}, which reduce to:
\begin{equation}
	[L_n, \mathcal{I}_K] = [\bar{L}_n, \mathcal{I}_K] = 0.
	\label{eq:commutation_full}
\end{equation}
Consequently, correlation functions depend only on the homotopy class of the contour along which $\mathcal{I}_K$ is inserted.

\subsection{Topological Defects in Symmetric Product Orbifolds}
\label{sec:orbifold-defects}

We now specialize to symmetric product orbifolds $\text{Sym}^N(\mathcal{M}) \equiv \mathcal{M}^{\otimes N}/S_N$, where $\mathcal{M}$ is a seed CFT of central charge $c$, and $S_N$ is the symmetric group of order $N!$. The total central charge is $c_{\text{total}} = Nc$. Following \cite{Gutperle:2024vyp, Knighton:2024noc, Gutperle:2024rvo}, we review two families of topological defects in these orbifolds: \emph{universal defects} (independent of seed theory details) and \emph{non-universal} or ``maximally fractional'' defects (constructed from seed theory data). Parallel constructions exist for boundary states \cite{Gaberdiel:2021kkp}.

\subsubsection{Symmetric product orbifold}
By definition, a symmetric product orbifold $\text{Sym}^N(\mathcal{M})$ is constructed by taking  $N$ copies of the seed CFT $\mathcal{M}$ and gauging the  $S_{N}$ permutation symmetry.

The Hilbert space of the symmetric product orbifold $\mathrm{Sym}^N(X)$ decomposes into twisted sectors, each labeled by a conjugacy class $[g]$ of the symmetric group $S_N$. Concretely, the decomposition takes the form
\begin{equation}\label{eq:hilbert-decomp}
	\mathcal{H} = \bigoplus_{[g]} \mathcal{H}_{[g]},
\end{equation}
%where the sum runs over all conjugacy classes $[g]$ of $S_N$. %, and $\mathrm{Inv}_{S_N}$ denotes the projection onto $S_N$-invariant states within each twisted sector $\mathcal{H}_{[g]}$.
where $[g]$ denotes a conjugacy class of $S_N$. Each class is characterized by a cycle structure
\begin{equation}
	[g] = 1^{l_1} 2^{l_2} \cdots N^{l_N}, \quad 
	\sum_{k=1}^N k\, l_k = N,
	\label{eq:cycle-structure}
\end{equation}
with $l_k$ counting cycles of length $k$. The untwisted sector ($[g]=[1]$) consists of $S_N$-invariant combinations of product states from $\mathcal{M}^{\otimes N}$, e.g.,
\begin{equation}
	\sum_{I=1}^N \mathcal{O}_I, \quad 
	\sum_{I,J=1}^N \mathcal{O}_I \mathcal{O}_J, \quad \dots
\end{equation}
where $\mathcal{O}_I$ denotes an operator in the $I$-th copy.

Twisted sector operators $\sigma_g$ (with $g\in S_N$) implement boundary conditions
\begin{equation}
	\mathcal{O}_I(e^{2\pi i} z) \, \sigma_g(0) = \mathcal{O}_{g(I)}(z) \, \sigma_g(0).
	\label{eq:twist-condition}
\end{equation}
Gauge-invariant twisted operators are obtained by averaging over conjugacy classes:
\begin{equation}
	\sigma_{[g]} = \frac{1}{|C(g)|} \sum_{h\in S_N} \sigma_{h^{-1} g h},
	\label{eq:gauge-invariant-twist}
\end{equation}
where $C(g)$ is the centralizer of $g$.

The partition function of the symmetric product orbifold is constructed by inserting the projection operator $P = \frac{1}{N!} \sum_{h \in S_N}h$ into the trace over each twisted sector Hilbert space \(\mathcal{H}_{[g]}\), thereby selecting only states invariant under the full symmetric group \(S_N\).

Geometrically, this corresponds to computing the partition function on a torus where the fields are twisted by a permutation \(g\) along the spatial cycle and by a permutation \(h\) along the temporal cycle. Consistency of the boundary conditions requires \([g,h]=0\); consequently, the sum is restricted to commuting pairs:
The torus partition function exhibits a standard orbifold form \cite{Klemm:1990df}:
\begin{equation}
	Z(\tau,\bar{\tau}) = 
	\frac{1}{N!} \sum_{\substack{g,h\in S_N \\ gh=hg}} 
	\farsquare{h}{g} \; (\tau,\bar \tau),
	\label{eq:orbifold-partition}
\end{equation}
where
\begin{equation}
	\farsquare{h}{g} \; (\tau,\bar \tau) = {\rm tr}_{{\cal H}_g} \Big( h \, e^{2\pi i \tau (L_0-\frac{c}{24}) } e^{-2\pi i \bar \tau (\bar L_0-\frac{c}{24}) } \Big).
	\label{eq:zhg}
\end{equation}
Here, the summation over \(g\) labels the distinct twisted sectors, while the summation over \(h\) enforces the projection onto \(S_N\)-invariant states.

Modular invariance of partition function (\ref{eq:orbifold-partition}) follows from  the modular transformation $\tau\to -1/\tau$ property of torus partition function in the individual sectors as:
\begin{equation}
	\farsquare{h}{g} (\tau, \bar \tau) \to  \farsquare{g^{-1}}{h} \Big(-\frac{1}{\tau},-\frac{1}{\bar \tau}\Big).
	\label{eq:modular_transform}
\end{equation}

\subsubsection{Universal Defects}
\label{subsubsec:universal-defects}

Via the folding trick~\cite{Oshikawa:1996dj, Bachas:2001vj}, topological defects can be realized as Ishibashi boundary states~\cite{Ishibashi:1988kg, Cardy:1989ir}, which in turn characterize $\mathbb{Z}_2$ permutation branes~\cite{Recknagel:2002qq, Drukker:2010jp, Cordova:2023qei}.
 Consider the doubled theory $\text{Sym}^N(\mathcal{M}) \times \text{Sym}^N(\mathcal{M})$, and let $R$ be an irreducible representation of $S_N$ with character $\chi_R$. The boundary state
\begin{equation}
	|B_R\rangle\!\rangle = 
	\sum_{[g]} \chi_R([g]) 
	\left( \sum_{n \in \mathcal{H}_{[g]}} |n\rangle_{(1),[g]} \otimes \overline{|n\rangle}_{(2),[g]} \right)
	\left( \sum_{m \in \mathcal{H}_{[g]}} |m\rangle_{(2),[g]} \otimes \overline{|m\rangle}_{(1),[g]} \right)
	\label{eq:universal-boundary-state}
\end{equation}
satisfies the topological gluing conditions
\begin{equation}
	\left( L_n^{(1)} - \bar{L}_{-n}^{(2)} \right) |B_R\rangle\!\rangle = 
	\left( L_n^{(2)} - \bar{L}_{-n}^{(1)} \right) |B_R\rangle\!\rangle = 0,
	\label{eq:top-gluing}
\end{equation}
where superscripts $(1),(2)$ label the two copies. This state defines an Ishibashi-like boundary state \cite{Ishibashi:1988kg, Cardy:1989ir} in the folded theory.

Unfolding back to a single copy yields the defect operator
\begin{equation}
	\mathcal{I}_R = \sum_{[g]} \chi_R([g]) \, P_{[g]} \otimes \bar{P}_{[g]},
	\label{eq:universal-defect}
\end{equation}
where $P_{[g]}$ is the identity projector on $\mathcal{H}_{[g]}$. From \eqref{eq:top-gluing}, it follows immediately that $\mathcal{I}_R$ is topological:
\begin{equation}
	[L_n, \mathcal{I}_R] = [\bar{L}_n, \mathcal{I}_R] = 0.
\end{equation}
Since $P_{[g]}$ is proportional to the identity in each twisted sector, $\mathcal{I}_R$ depends only on the orbifold group $S_N$ and not on detailed data of $\mathcal{M}$.

\subsubsection{Non-universal (Maximally Fractional) Defects}
\label{subsec:nonuniversal-defects}

The universal defects discussed in the previous section are constructed as projectors proportional to the identity operator in each twisted sector. Consequently, they are independent of the detailed structure of the seed CFT $\mathcal{M}$; they can be viewed as being built from the trivial (identity) defect in the seed theory.

To construct defects that probe the internal structure of $\mathcal{M}$, we employ a ``maximally fractional'' approach analogous to that used for D-branes in symmetric orbifolds \cite{Gaberdiel:2021kkp}. Let $K$ denote a topological interface in the seed theory $\mathcal{M}$, with expansion coefficients $d_{K\mathbf{i}}$ as in \eqref{eq:interface_operator}.

For a twisted sector corresponding to a single $w$-cycle, we define the defect operator
\begin{equation}
	\mathcal{I}_{K(w)} = \sum_{\mathbf{i}} d_{K\mathbf{i}} \, P_{(w)}^{\mathbf{i}},
	\label{eq:single-cycle-defect}
\end{equation}
where $P_{(w)}^{\mathbf{i}}$ projects onto specific states in the $w$-cycle sector:
\begin{equation}
	P_{(w)}^{(i,\bar{j};\alpha,\beta)} = 
	\sum_{\mathbf{n},\bar{\mathbf{n}}} 
	\bigl( |i,\mathbf{n}\rangle \otimes |\bar{j},\bar{\mathbf{n}}\rangle \bigr)_{(w)}^{(\alpha)}
	\bigl( \langle i,\mathbf{n}| \otimes \langle \bar{j},\bar{\mathbf{n}}| \bigr)_{(w)}^{(\beta)}.
	\label{eq:w-cycle-projector}
\end{equation}
Here $i,\bar{j}$ label left- and right-moving representations in $\mathcal{M}$, $\mathbf{n},\bar{\mathbf{n}}$ denote descendant states, and $\alpha,\beta$ index possible multiplicity labels.

The full orbifold defect is constructed by assembling these building blocks across all cycles in a given conjugacy class. For a conjugacy class $[g]$ with cycle structure \eqref{eq:cycle-structure}, we define
\begin{equation}
	\mathcal{I}_{K,R} = \sum_{[g]} \chi_R([g]) \prod_{j=1}^{N} \prod_{k_j=1}^{l_j} \mathcal{I}_{K(j)},
	\label{eq:full-nonuniversal-defect}
\end{equation}
where $\mathcal{I}_{K(j)}$ denotes the defect operator \eqref{eq:single-cycle-defect} for a $j$-cycle.

\paragraph{Special Case: Rational CFT and Verlinde Lines}
When the seed theory $\mathcal{M}$ is a diagonal rational CFT, a particularly important class of defects arises from the Verlinde lines \cite{Verlinde:1988sn}. For a primary label $a$ of $\mathcal{M}$, the corresponding Verlinde line is given by \cite{Petkova:2000ip,Petkova:2001ag}
\begin{equation}
	\mathcal{I}_a = \sum_i \frac{S_{ai}}{S_{0i}} P^{i\bar{i}}.
	\label{eq:verlinde-line}
\end{equation}
In the tensor product theory $\mathcal{M}^{\otimes N}$, we use the same defect label $a$ for all factors to construct an $S_N$-invariant ``maximally fractional'' defect.

For a single $w$-cycle twisted sector, the doubled boundary state takes the form
\begin{equation}
	|B_a\rangle\!\rangle_{(w)} = \sum_i \frac{S_{ai}}{S_{0i}} \sum_{n} |i, n\rangle^{(1)}_{(w)} \otimes \overline{|i,n\rangle}^{(2)}_{(w)} \sum_{m} |i,m\rangle^{(2)}_{(w)} \otimes \overline{|i,m\rangle}^{(1)}_{(w)},
	\label{eq:doubled-boundary-state}
\end{equation}
where the sums over $n,m$ include all descendants (including fractional Virasoro modes) of the $w$-twisted sector primary $|i\rangle_{(w)}$.

Assembling these cycle-wise boundary states with $S_N$ representation data yields the full boundary state
\begin{equation}
	|B_a^R\rangle\!\rangle = \sum_{[g] \in S_N} \chi_R([g]) \prod_{j=1}^N \prod_{k_j=1}^{l_j} |B_a\rangle\!\rangle_{(j)},
	\label{eq:full-boundary-state}
\end{equation}
where the product runs over all cycles in the conjugacy class $[g]$ as specified in \eqref{eq:cycle-structure}. Equivalently, in the unfolded picture, the corresponding topological defect operator is
\begin{equation}
	\mathcal{I}_{R}^{a} = \sum_{[g]} \chi_R([g]) \prod_{j=1}^{N} \prod_{k_j=1}^{l_j} \mathcal{I}_{(j)}^{a},
	\label{eq:rational-orbifold-defect-final}
\end{equation}
where $\mathcal{I}_{(j)}^{a}$ is the $j$-cycle projector
\begin{equation}
	\mathcal{I}_{(j)}^{a} = \sum_i \frac{S_{ai}}{S_{0i}} P_{(j)}^{i\bar{i}}, \qquad
	P_{(j)}^{i\bar{i}} = \sum_{n,\bar{n}} |i,n\rangle_{(j)} \otimes \overline{|i,n\rangle}_{(j)} \, \langle i,n|_{(j)} \otimes \overline{\langle i,n|}_{(j)}.
	\label{eq:cycle-verlinde-projector}
\end{equation}

It has been shown in \cite{Gutperle:2024vyp} that these boundary states satisfy the Cardy conditions, thereby ensuring they define consistent topological defects in the symmetric orbifold. These non-universal defects depend explicitly on the modular data of the seed CFT and establish a deep correspondence between the categorical symmetry of $\mathcal{M}$ and that of its symmetric product $\mathrm{Sym}^N(\mathcal{M})$.

\section{Defect relative entropy and topological defects}
\label{sec:orbifold-defect relative}
	
In this section, we compute the relative entropy for both universal and non-universal topological defects introduced in the previous section. We begin by reviewing the notion of defect relative entropy in two-dimensional conformal field theories, and then proceed to evaluate it in the setting of symmetric product orbifolds.

\subsection{Defect relative entropy in 2d CFT }

To compute the relative entropy between two topological interfaces:
\[
\mathcal{I}_K = \sum_{\mathbf{i}} d_{K\mathbf{i}} \, P^{\mathbf{i}} 
\quad \text{and} \quad 
\mathcal{I}_{K'} = \sum_{\mathbf{i}} d_{K'\mathbf{i}} \, P^{\mathbf{i}},
\]
we employ the replica trick (\ref{eq:replica}). The type of topological defect is encoded in the expansion coefficients \(d_{K\mathbf{i}}\) and \(d_{K'\mathbf{i}}\).
The key quantity in this approach is 
\[
\operatorname{Tr}\!\big(\rho_K \, \rho_{K'}^{\,n-1}\big),
\]
which for integer $n$ corresponds to a generalized partition function (or correlation function) on an $n$-sheeted Riemann surface. This construction explicitly breaks the $\mathbb{Z}_n$ replica symmetry \cite{Lashkari:2015dia}. Thus,  computing the relative entropy reduces to computing the partition function via the  replica trick Eq. (\ref{eq:replica}) on an $n$-sheeted Riemann surface with $2n$ operator insertions.

Specifically, the $n$-th replica partition function is given by a torus correlator with $2n$ insertions of the interface operators \cite{Sakai:2008tt, Brehm:2015lja, Brehm:2015plf, Gutperle:2015kmw}. More precisely, the replica partition function for calculating the relative entropy across a topological interface takes the form of a torus partition function with $2n$ interface operator insertions as:
\begin{equation}
	\begin{aligned}
		Z_n(K, K') &= \operatorname{tr} \left[ \left( \cI_K e^{-t H} I_K^\dagger e^{-t H} \right)\left( \cI_{K'} e^{-t H} I_{K'}^\dagger e^{-t H} \right)^{(n-1)} \right] \\
		&= \operatorname{tr} \left[\left( \cI_K \cI_K^\dagger \right) \left( \cI_{K'} \cI_{K'}^\dagger \right)^{(n-1)} e^{-2n t H} \right] \\
		&= \sum_{(i,\bar{j})} \operatorname{Tr} \left[(d_{K\mathbf{i}} d_{K\mathbf{i}}^*) (d_{K'\mathbf{i}} d_{K'\mathbf{i}}^*)^{(n-1)} \right] \, \chi_i\!\left(e^{-2n t}\right) \, \chi_{\bar{j}}\!\left(e^{-2n t}\right) .
		\label{eq:dureplica_partition}
	\end{aligned}
\end{equation}
and
\begin{equation}
	\begin{aligned}
		Z_n(K) &= \operatorname{tr} \left[ \left( \cI_K e^{-t H} \cI_K^\dagger e^{-t H} \right)^n \right] \\
		&= \operatorname{tr} \left[ \left( \cI_K \cI_K^\dagger \right)^n e^{-2n t H} \right] \\
		&= \sum_{(i,\bar{j})} \operatorname{Tr} \left[ (d_{K\mathbf{i}} d_{K\mathbf{i}}^*)^n \right] \, \chi_i\!\left(e^{-2n t}\right) \, \chi_{\bar{j}}\!\left(e^{-2n t}\right) .
		\label{eq:replica_partition}
	\end{aligned}
\end{equation}
Here, $H$ denotes the Hamiltonian on a cylinder:
\begin{equation}
	\begin{aligned}
H = L_0 + \bar{L}_0 - c/12 .
	\end{aligned}
\end{equation}
  We employed the commutation relation \eqref{eq:commutation_full} in the equations above. Here, $\chi_i$ denotes the character in the representation $\mathcal{V}_i$, and the parameter $t$ is related to the UV and IR cutoffs $\epsilon$ and $L$ via
\begin{equation}
	\begin{aligned}
	t = \frac{2\pi^2}{\log(L/\epsilon)} .
	\end{aligned}
\end{equation}
The trace \(\operatorname{Tr}\) is taken over the multiplicity indices \(\alpha, \beta\), treating \(d_{K(i,\bar{j}; \alpha, \beta)}\) as a matrix, with \(d_{K\mathbf{i}}^* \equiv d_{K(i,\bar{j}; \beta, \alpha)}^*\).

Using the modular transformation properties of the characters \(\chi_i\), the entanglement entropy \(S(\CI_K) \) is derived from the analytic continuation of \(\log Z_n(K)\) as \(n \to 1\) \cite{Brehm:2015plf}:
\begin{equation}
	S(\CI_K) = \frac{c}{6} \log \left( \frac{L}{\epsilon} \right) - \sum_{(i,\bar{j})} \operatorname{Tr} \left[ p^K_{\mathbf{i}} \log \left( \frac{p^K_{\mathbf{i}}}{p^{\text{Id}}_{\mathbf{i}}} \right) \right] .
	\label{eq:entropy_general}
\end{equation}
 The defect relative entropy is derived as \cite{Ghasemi:2026hoa}:
\begin{align}\label{KUL-DRE1}
	D(\cI_K\|\cI_{K'})& = -\, \partial_n \log G_{n}(K|K')\Big|_{n \to 1} \nonumber\\&
	=\sum_{(i, \bar j)}\,\Tr\left[ p^K_{\bf i}\,\log  \frac{p^K_{\bf i}}{p^{K'}_{\bf i}}\right] \ ,
\end{align}
where $G_{n}(K|K')$ defined as:
\begin{align}\label{Parti-gen1}
	G_{n}(K|K')&=\frac{	Z_{n}(K,K')(Z_{1}(K))^{n-1} }{	Z_{n}(K) (Z_{1}(K'))^{n-1}}. 
	%	\nonumber\\&
	%	=\frac{[\mathcal{F}(1,\alpha)]^{(n-1)}[\mathcal{G}(n,\alpha,\beta)]}{[\mathcal{F}(1,\beta)]^{(n-1)}[\mathcal{F}(n,\alpha)]},
\end{align}
The probability distributions \(p^K\) and \(p^{\text{Id}}\) are defined via the modular \(S\)-matrix:
\begin{equation}
	p^K_{\mathbf{i}} \equiv \frac{S_{0i} (S_{0\bar{j}})^*}{\sum_{(i,\bar{j})} S_{0i} (S_{0\bar{j}})^* \operatorname{Tr}[d_{K\mathbf{i}} d_{K\mathbf{i}}^*]} \, d_{K\mathbf{i}} d_{K\mathbf{i}}^*, \qquad
	p^{\text{Id}}_{\mathbf{i}} \equiv S_{0i} (S_{0\bar{j}})^* \delta_{\alpha\beta}.
	\label{eq:probability_distributions}
\end{equation}
The expression~(\ref{KUL-DRE1}) can be interpreted as the Kullback--Leibler (KL) divergence~\cite{KullbackLeibler} between the probability distribution associated with the defect operator $\CI_K$  and that associated with the defect $\CI_K'$ on the $\mathrm{CFT}_1$ side. 

In a similar manner, we can derive an expression for the defect sandwiched Renyi relative entropy \cite{Ghasemi:2026hoa}:
\begin{align}\label{SNRE-2}
	&D_{n}(\cI_K\|\cI_{K'})
	=\frac{1}{1-n}\log\sum\limits_{(i,\bar j)}\Tr\left[ (p^K_{\bf i})^{n}\, (p^{K'}_{\bf i})^{(1-n)}  \right] 
	.  
\end{align}
The expression (\ref{SNRE-2}) for the special value of $n=\frac{1}{2}$ is
related to the fidelity, which is a natural generalization of the notion of pure states overlap, $\arrowvert \langle\phi|\psi\rangle\arrowvert$. Quantum fidelity can be used as a proper tool in characterizing quantum phase  transition \cite{Shi-Gu:2008zq}. Defect fidelity was obtained as \cite{Ghasemi:2026hoa}:	
\begin{align}\label{FID-1}
	F(\cI_K\|\cI_{K'})
	=\sum\limits_{(i,\bar j)}\Tr (\sqrt{p^K_{\bf i}}\,\sqrt{p^{K'}_{\bf i}})
	.  
\end{align}
It reduces to the fidelity between two probability distributions.

\subsubsection{Diagonal Theories}
For diagonal theories, where \(\cM_{i\bar{j}} = \delta_{i\bar{j}}\), the theory is the same on both sides. Topological interfaces are in one-to-one correspondence with primary operators labeled by \(a\) \cite{Petkova:2000ip}:
\begin{equation}
	\cI_a = \sum_i \frac{S_{ai}}{S_{0i}} P^i, \qquad
	P^i \equiv \sum_{n,\bar{n}} |i, n\rangle \otimes |i, \bar{n}\rangle \langle i, n| \otimes \langle i, \bar{n}|.
	\label{eq:diagonal_interface}
\end{equation}
In this case, the probability distributions simplify to:
\begin{equation}
	p^a_i = |S_{ai}|^2, \qquad p^{\text{Id}}_i = |S_{0i}|^2,
	\label{eq:probabilities_diagonal}
\end{equation}
and the entanglement entropy becomes \cite{Brehm:2015plf,Gutperle:2015kmw}:
\begin{equation}
	S(\cI_a) = \frac{c}{6} \log \left( \frac{L}{\epsilon} \right) - 2 \sum_i |S_{ai}|^2 \log \left| \frac{S_{ai}}{S_{0i}} \right| .
	\label{eq:entropy_diagonal}
\end{equation}
The first term represents the universal area law contribution, while the second term is a constant topological term. The defect relative entropy is given by \cite{Ghasemi:2026hoa}:
\begin{equation}
	D(\cI_a\|\cI_{a'}) = \sum_i |S_{ai}|^2 \log \left| \frac{S_{ai}}{S_{a'i}} \right|^{2} .
	\label{DRE_diagonal}
\end{equation}
Similarly, we can obtain the \textit{ sandwiched R\'enyi defect relative entropy} and the \textit{defect fidelity} as \cite{Ghasemi:2026hoa}. The quantum fidelity  then becomes: 
\begin{equation}\label{DF_diag_RCFT}
	F(\CI_a\|\CI_{a'}) = \sum\limits_{j} \left|\mathcal{S}_{aj}\right|\, \left|\mathcal{S}_{a'j}\right|.
\end{equation} 
It is notable that relations (\ref{DRE_diagonal}) and (\ref{DF_diag_RCFT}) coincide with the left-right relative entropy and left-right fidelity, respectively, introduced in~\cite{Ghasemi:2024wcq}. This correspondence arises because the topological defects in a diagonal rational CFT are in one-to-one correspondence with boundary states sharing the same expansion coefficients.

%Moreover, we observe an isomorphism between the \emph{relative entanglement sector} defined in~\cite{Ghasemi:2024wcq} and the \emph{defect relative sector} \cite{Ghasemi:2026hoa}. By definition, the defect relative sector is comprised of topological defects with zero defect relative entropy.

\subsection{Defect relative entropy in symmetric product orbifold}
\label{sec:ee2}

In this subsection, we will use the formulas that were reviewed in the previous subsection to calculate the relative entropy between the topological defects as defined in previous.

\subsubsection{Universal defects}
%\subsection{}
\label{subsec4.1}
Recall that the universal defects are labeled by the choice of irreducible representation $R$ of $S_N$ with character $\chi_{R}([g])$ being the character of the conjugacy class $[g]$ in the representation $R$,  independent of the  details of the seed CFT. Therefore, we choose two defects in two representations $R$ and $R'$ with corresponding characters $\chi_{R}([g])$ and $\chi_{R'}([g])$, respectively. For universal topological defects,
\[
\mathcal{I}_R = \sum_{[g]} \chi_R([g]) \, P_{[g]} \otimes \bar{P}_{[g]}
\quad \text{and} \quad 
\mathcal{I}_{R'} = \sum_{[g]} \chi_{R'}([g]) \, P_{[g]} \otimes \bar{P}_{[g]},
\]
the corresponding torus partition functions are given by:
 \begin{align}
 	Z_{n}(K,K') &= {\rm tr} \Big[ ({\cal I}_K)^{2} ({\cal I}_{K'})^{2(n-1)} e^{- 2 n t H}\big] \nonumber \\
 	&= \sum_{[g]} \big[\chi_R([g]) \big]^{2} \big[\chi_{R'}([g]) \big]^{2(n-1)} \tr\big( P_{[g]} \bar P_{[g]}  e^{- 2 n t H}\big)~,
 \end{align}
 and
\begin{align}
Z_{n}(K) &= {\rm tr} \Big[  {\cal I}^{2n} e^{- 2 n t H}\big] \nonumber \\
&= \sum_{[g]} \big[\chi_R([g]) \big]^{2n} \tr\big( P_{[g]} \bar P_{[g]}  e^{- 2 n t H}\big)~,
\end{align}
where we used the reality of the characters which implies that ${\cal I }^\dagger ={\cal I}$. Here, the modular parameter is given by 
\begin{align}
	\tau =  i \frac{2\pi n}{   \ln \frac{L}{ \epsilon}}~,
\end{align}
The partition functions can be expressed as a sum over twisted sectors:
\begin{align}
Z_{n}(K,K') &= \frac{1}{|G|} \sum_{hg=gh} \big[\chi_R([g]) \big]^{2} \big[\chi_{R'}([g]) \big]^{2(n-1)}  \farsquare{h}{g}( \tau, \bar \tau)~,\\
	Z_{n}(K) &= \frac{1}{|G|} \sum_{hg=gh} \big[\chi_R([g]) \big]^{2n}   \farsquare{h}{g}( \tau, \bar \tau)~.
\end{align}
where the sum is taken over twisted sectors labeled by $g$ with a projection onto $S_N$-invariant states implemented by $h$. Here $h$ runs over the centralizer of $g$ (all elements commuting with $g$). Taking the cutoff $L \to \infty$ sends $\tau \to 0$, which requires performing a modular transformation on the torus partition function,
\begin{align} 
	Z_{n}(K,K') &= \frac {1}{ |G|} \sum_g \sum_{h\in C[g]} \big[\chi_R([g])\big]^{2} \big[\chi_{R'}([g]) \big]^{2(n-1)}  \farsquare{g^{-1} }{h }( - \frac{1}{ \tau} ,-\frac{1}{ \bar \tau})~,\\
	Z_{n}(K) &= \frac {1}{ |G|} \sum_g \sum_{h\in C[g]} \big[\chi_R([g]) \big]^{2n}  \farsquare{g^{-1} }{h }( - \frac{1}{ \tau} ,-\frac{1}{ \bar \tau})~.
\end{align}
Taking the limit $\tau \to 0$ implies $-\frac{1}{\tau} \to \infty$ for the modular-transformed partition function. Hence, the partition function is dominated by the sector containing the primary with the smallest conformal dimension. For the symmetric orbifold, this corresponds to the untwisted vacuum sector.
Furthermore, note that in the centralizer $C(g)$ of any element $g \in S_N$, the identity element is always present, since it commutes trivially with any group element. Consequently, in the limit $\tau \to 0$, the partition function is dominated by the untwisted sector.
\begin{align}
	\lim_{\tau \to 0}  Z_{n}(K,K') 
	&\sim\lim_{\tau \to 0}   \sum_{g\in S_N} \frac{1}{|G|} \big[\chi_R([g])\big]^{2} \big[\chi_{R'}([g]) \big]^{2(n-1)}  \farsquare{g }{1 }( -\frac{1}{ \tau} ,-\frac{1}{ \bar \tau}) ~,\\
	\lim_{\tau \to 0}  Z_{n}(K) 
	&\sim\lim_{\tau \to 0}   \sum_{g\in S_N} \frac{1}{|G|}  \big[\chi_R([g]) \big]^{2n} \farsquare{g }{1 }( -\frac{1}{ \tau} ,-\frac{1}{ \bar \tau}) ~.
\end{align}
Note that in the torus partition function,  we replace $g^{-1}$ by $g$ due to the reality of the character. In the limit $\tau \to 0$ each term in the partition function contributes the same leading term, which is independent of $g$.
To compute the partition functions above, we first need to evaluate  $\farsquare{g }{1 }  (-\frac{1}{\tau}, -\frac{1}{ \bar \tau} ) $.

Thus, to compute this quantity we proceed as follows:   A given element $g\in S_N$  is represented by $k$ cycles of lengths $n_1,n_2, \cdots n_k$ satisfying $\sum_{i=1}^k n_i = N$. In the tensor product theory, the partition function with a $g$-insertion is given by
\begin{align}
	\farsquare{g }{1 }  (-\frac{1}{\tau}, -\frac{1}{ \bar \tau} ) = \prod_{i=1}^k  Z_{seed}(-  n_i \frac{1}{ \tau}  , -n_i \frac{1}{ \bar\tau})~.
\end{align}
In the limit $\tau \to 0 $  the partition function of the seed theory is dominated by the vacuum contribution 
\begin{align}
	\lim_{\tau \to 0} \farsquare{g }{1 }  (-\frac{1}{ \tau}, -\frac{1}{ \bar \tau} ) &\sim  \prod_{i=1}^k  e^{- \frac{2\pi i }{ \tau }  n_i c_{seed}/24 } e^{\frac{2\pi i }{ \bar  \tau }  n_i c_{seed}/24 } \nonumber \\
	&\sim e^{-\frac{2\pi i}{  \tau }  N c_{seed}/24 } e^{\frac{2\pi i}{ \bar  \tau }  N c_{seed}/24 } \nonumber \\
	&\sim   \exp\Big( {\frac{c}{ 12 } \frac {1}{ K} \ln \frac{L}{ \epsilon} }\Big) ~,
\end{align}
 where the central charge is $c= N c_{seed}$. 
 Using Eq.(\ref{eq:replica}),  we obtain 
\begin{align}\label{Parti-gen-de}
	G_{n}(K|K')&=\frac{	Z_{n}(K,K')(Z_{1}(K))^{n-1} }{	Z_{n}(K) (Z_{1}(K'))^{n-1}} 
	\nonumber\\&
	\sim\lim_{\tau \to 0}\frac{\sum_{g\in S_N} \frac{1}{|G|} \big[\chi_R([g])\big]^{2} \big[\chi_{R'}([g]) \big]^{2(n-1)} \,\big[\sum_{g\in S_N}\frac{1}{|G|} \big[\chi_R([g])\big]^{2}]^{n-1}}{\sum_{g\in S_N} \frac{1}{|G|}  \big[\chi_R([g]) \big]^{2n}[\sum_{g\in S_N} \frac{1}{|G|} \, \big[\chi_{R'}([g]) \big]^{2}]^{n-1}},
\end{align}
where the sum over $g$ is implicitly understood as a sum over conjugacy classes $[g]$.
From this we arrive at the final expression for the relative entropy between two universal topological defects:
\begin{align}
	D(\cI_K\|\cI_{K'})
	=-\partial_{n}G_{n}(K|K')=
	 \sum_g  \frac{1}{ |G|}  \big[\chi_R([g]) \big]^2 \ln  \Big(\frac{[ \chi_R([g])]^2}{[ \chi_{R'}([g])]^2}\Big) \ .
\end{align}
This can be further simplified to
\begin{align}\label{result-univ}
	D(\mathcal{I}_K \| \mathcal{I}_{K'})
	= \sum_{[g]} p_{R}([g]) \, \ln \frac{p_{R}([g])}{p_{R'}([g])},
\end{align}
where $p_{R}([g])$ is a probability distribution defined in terms of the character of the representation:
\begin{align}\label{Prob-CHA}
	p_{R}([g]) = \frac{1}{|G|} \, \chi_{R}([g])^2 .
\end{align}
In deriving this result, we have used both the orthogonality relation of characters and the fact that characters of $S_N$ are real:
\begin{align}\label{charorth}
	\frac{1}{|G|} \sum_{[g]} \chi_{R}([g])^* \chi_{S}([g]) = \delta_{RS}.
\end{align}
%Expression~(\ref{result-univ}) can be interpreted as the Kullback--Leibler (KL) divergence~\cite{KullbackLeibler} between probability distributions defined by the characters of $S_N$.
Expression~(\ref{result-univ}) admits a natural interpretation as the Kullback--Leibler (KL) divergence~\cite{KullbackLeibler} between two probability distributions defined on the set of conjugacy classes of $S_N$. Specifically, the squared characters $\chi_R([g])^2$ (appropriately normalized) define a probability measure, and the relative entropy between two such measures reproduces precisely the form of the KL divergence.

The defect fidelity is given by
\begin{equation}\label{DF_UNIDEF}
	F(\mathcal{I}_a \| \mathcal{I}_{a'}) = \sum_{[g]} \sqrt{p_{R}([g])} \sqrt{p_{R'}([g])},
\end{equation}
which reduces to the fidelity between two probability distributions defined in Eq.~\eqref{Prob-CHA}.

As can be seen from Eqs.~\eqref{result-univ} and \eqref{DF_UNIDEF}, for universal defects the characters of the permutation group $S_N$ play the role of a probability distribution.

 \subsubsection{Non-universal defects}   
%\subsection{}
To calculate the defect relative entropy in the case of the non-universal defect, we replace the topological defects with (\ref{eq:full-nonuniversal-defect}) or (\ref{eq:rational-orbifold-defect-final}), and calculate the corresponding partition functions. The computations follow the same strategy as the previous subsection,  with the sum  taken over the centralizer of $g$. The leading contribution is obtained from the partition function in the $h=1$ sector, before modular transformation, which is then mapped to the dominant untwisted sector after modular transformation as:
 \begin{align}
 	Z_n(K,K')  &= \sum_{g} \frac{1}{|G|}\big(\chi_{R}([g])\chi\dag_{R}([g])\big) \big(\chi_{R'}([g])\chi\dag_{R'}([g])\big)^{(n-1)} \prod_{j=1}^N \left( \sum_i  \left| \frac{S_{ai}}{ S_{0i}}\right|^{2} \left| \frac{S_{a'i}}{ S_{0i}}\right|^{2(n-1)}  \chi_i (\frac{\tau}{ j})  \bar \chi_i (\frac{\bar \tau}{ j}) \right)^{l_j}+\cdots~.\\
 	 Z_n(K) & = \sum_{g} \frac{1}{|G|} \big(\chi_R([g])\chi\dag_{R}([g])\big)^{n} \prod_{j=1}^N \left( \sum_i \left| \frac{S_{ai}}{ S_{0i}}\right|^{2n}  \chi_i (\frac{\tau}{ j})  \bar \chi_i (\frac{\bar \tau}{ j}) \right)^{l_j}+\cdots~.
 \end{align}
The limit $\tau\to 0$ of these partition functions are evaluated through an  S-modular transformation by transforming the RCFT characters. This process  leads to:
\begin{align}\label{znonuni01}
	\lim_{\tau \to 0}  Z_n(K,K') 
	&\sim\lim_{\tau \to 0} \sum_g \frac{1}{ |G|} \big(\chi_R([g])\big)^{2}\big(\chi_{R'}([g])\big)^{2(n-1)}\\
\nonumber
	 &\times\prod_{j=1}^N \left( \sum_{i,k,\bar k} \left| \frac{S_{ai}}{ S_{0i}}\right|^{2} \left| \frac{S_{a'i}}{ S_{0i}}\right|^{2(n-1)}  S_{ik} \; S^*_{i \bar k} \; \chi_k (-j \frac{1}{ \tau}) \bar \chi_{\bar k} (-j \frac{1}{ \bar \tau}) \right)^{l_j}+\cdots~.
\end{align}
\begin{align}\label{znonuni1}
    \lim_{\tau \to 0}  Z_n(K) 
&\sim\lim_{\tau \to 0} \sum_g \frac{1}{ |G|} \big(\chi_R([g])\big)^{2n} \prod_{j=1}^N \left( \sum_{i,k,\bar k} \left| \frac{S_{ai}}{ S_{0i}}\right|^{2n}  S_{ik} \; S^*_{i \bar k} \; \chi_k (-j \frac{1}{ \tau}) \bar \chi_{\bar k} (-j \frac{1}{ \bar \tau}) \right)^{l_j}+\cdots~.
\end{align}
In the $\tau \to 0$ limit, the dominant contributions to (\ref{znonuni01}) and (\ref{znonuni1}) come from the vacuum character $\chi_0$, which has the asymptotic behavior  
\[
\chi_0 \sim q^{-\frac{c_{\mathrm{seed}}}{24}} \, \bar q^{-\frac{c_{\mathrm{seed}}}{24}} \qquad (q \to 0),
\]
where $q = e^{2\pi i \tau}$. Using this expansion together with the relation $\sum_j \ell_j = N$ for all permutations $g \in S_N$, we obtain
\begin{align}
	\lim_{\tau \to 0}  Z_n(K,K') 
	&\sim   \sum_g \frac{1}{|G|}\big(\chi_R([g])\big)\big(\chi_{R'}([g])\big)^{2(n-1)} \Big( \sum_i|S_{ai}|^{2} |S_{a'i}| ^{2(n-1)} |S_{0i}|^{2-2n}\Big)^N \exp\big( {\frac{c}{ 12 } \frac{1}{ n} \ln \frac{L}{ \epsilon} }\big) +\cdots~.
\end{align}
\begin{align}
   \lim_{\tau \to 0}  Z_n(K) 
&\sim   \sum_g \frac{1}{|G|} \big(\chi_R([g])\big)^{2n} \Big( \sum_i |S_{ai}| ^{2n} |S_{0i}|^{2-2n}\Big)^N \exp\big( {\frac{c}{ 12 } \frac{1}{ n} \ln \frac{L}{ \epsilon} }\big) +\cdots~.
\end{align}
 Using Eq.\ref{eq:replica},  we then obtain 
\begin{align}\label{Parti-gen-de2}
	G_{n}(K|K')&=\frac{	Z_{n}(K,K')(Z_{1}(K))^{n-1} }{	Z_{n}(K) (Z_{1}(K'))^{n-1}} 
	\nonumber\\&
	\sim\lim_{\tau \to 0}\frac{ \sum_g \frac{1}{|G|}\big(\chi_R([g])\big)^{2}\big(\chi_{R'}([g])\big)^{2(n-1)} \Big( \sum_i|S_{ai}|^{2} |S_{a'i}| ^{2(n-1)} |S_{0i}|^{2-2n}\Big)^N \,}{\sum_g \frac{1}{|G|} \big(\chi_R([g])\big)^{2n} \Big( \sum_i |S_{ai}| ^{2n} |S_{0i}|^{2-2n}\Big)^N}
%&\\ \times \frac{\big[\sum_g \frac{1}{|G|} \big(\chi_R([g])\big)^{2n} \Big( \sum_i |S_{ai}| ^{2n} |S_{0i}|^{2-2n}\Big)^N\big]^{n-1}}{ \big[\sum_g \frac{1}{|G|} \big(\chi_R([g])\big)^{2n} \Big( \sum_i |S_{ai}| ^{2n} |S_{0i}|^{2-2n}\Big)^N\big]^{n-1}},
\end{align}
where the sum over $g$ is implicitly understood as a sum over conjugacy classes $[g]$. Taking the logarithm of this expression results in the defect relative entropy:
\begin{align}
	D(\mathcal{I}_{a} \| \mathcal{I}_{a'}) =  \sum_g  \frac{1}{ |G|}  \big[\chi_{R}([g]) \big]^2 \ln  \Big(\frac{ \big[\chi_{R}([g]) \big]^2}{ \big[\chi_{R'}([g]) \big]^2}\Big)+ N \sum_{i} |S_{ai}|^2 \ln \left(\left| \frac{S_{ai}}{  S_{a'i}}\right|^2\right)~,
\end{align}
To obtain the above expression, we used the unitarity and symmetry of the modular $S$ matrix, in particular $\sum_j |S_{aj}|^2=1$. This expression can be rewritten as:
\begin{align}\label{DRE-DIAG}
	D(\mathcal{I}_{a} \| \mathcal{I}_{a'})=  \sum_{g}\, p_{R}(g)\,\ln  \frac{p_{R}(g)}{p_{R'}(g)}+ N \sum_{i} p_{ai} \ln \left( \frac{p_{ai}}{  p_{a'i}}\right)~,
\end{align}
where $p_{ai}=|S_{ai}|^2$. As we can see, the defect relative entropy between non-universal defects, reduces to two kind of  Kullback--Leibler divergences, in which the characters of the permutation group $S_{N}$ and the modular $S$ matrix take on the role of a probability distribution.
%It is interesting that for the ``maximally fractional" universal defect, the defect relative entropy is the sum of the contribution of the universal defect  (\ref{result-univ}) and $N$ times the contribution of the RCFT topological defect \cite{Ghasemi:2026hoa}. This indicates that the maximally fractional defect can be viewed, in some sense,  as a product of the RCFT and symmetric orbifold defect.
For a defect described in Eq. \ref{eq:full-nonuniversal-defect},  we can express Eq. \ref{DRE-DIAG}  as follows:
\begin{align}
	D(\mathcal{I}_{K}\| \mathcal{I}_{K'})& =  \sum_{g}\, p_{R}(g)\,\ln  \frac{p_{R}(g)}{p_{R'}(g)}+
	N\sum_{(i, \bar j)}\,\Tr\left[ p^K_{\bf i}\,\ln  \frac{p^K_{\bf i}}{p^{K'}_{\bf i}}\right] \ .
\end{align}
For the maximally fractional universal defect, the defect relative entropy splits into two terms: the contribution from the universal defect itself plus $N$ times the contribution from the RCFT topological defect~\cite{Ghasemi:2026hoa}. This suggests that the maximally fractional defect can be understood as a kind of product of the RCFT defect and the symmetric orbifold defect—a feature already observed in the defect entanglement entropy~\cite{Gutperle:2024rvo}.

It would be interesting to explore whether more general constructions—for instance, those based on D-branes in symmetric orbifolds~\cite{Belin:2021nck} or the extended defect frameworks in~\cite{Benjamin:2025knd}—are possible, and whether they yield a more intricate interplay between RCFT defects and universal defects.

\section{Discussion}
\label{sec:Discussion}

In this work, we have computed the defect relative entropy between topological defects in the symmetric product orbifold $\mathrm{Sym}^N(M)=M^{\otimes N}/S_N$. This quantity, introduced in~\cite{Ghasemi:2026hoa}, provides a measure of distinguishability in the space of topological defects. Our analysis focuses on two distinct classes: universal defects, which realize the $\mathrm{Rep}(S_N)$ non-invertible symmetry, and non-universal defects.

We find that the defect relative entropy reduces to a Kullback--Leibler divergence. The resulting expression decomposes naturally into two contributions: one governed by characters of $S_N$, and the other controlled by modular $S$-matrix elements of the seed RCFT. Remarkably, both structures enter as probability distributions, providing an information-theoretic interpretation of permutation group data and modular data within the symmetric orbifold.

The structure of the divergence depends sensitively on the defect class. For universal defects, only the permutation group data contributes. For maximally fractional defects, in contrast, both permutation and modular data contribute and together define the relevant probability distributions. This distinction reflects the underlying algebraic structure of the defects and highlights the rich interplay between the orbifold data and the seed CFT.

%Our results demonstrate that defect relative entropy serves as a sensitive probe of both the representation theory of $S_N$ and the modular data of the seed theory, unifying them within a single information-theoretic framework. More broadly, this work illustrates how entanglement-theoretic quantities can provide new insights into the structure of non-invertible symmetries and their realizations in tractable, interacting CFTs.

The simplification to a KL divergence appears to be a distinctive feature of topological defects. A natural direction for future work is to investigate whether analogous formulas exist for conformal (non-topological) defects that share structural properties with the universal topological defects studied here. Such an extension would be particularly significant from the perspective of AdS/CFT, where these defects are expected to correspond to finite-tension branes in the bulk~\cite{Gaberdiel:2021kkp, Belin:2021nck, Harris:2025wak, Harris:2025klp}.

It would also be worthwhile to explore more general constructions of defects in symmetric product orbifolds, for instance along the lines of the D-brane constructions in~\cite{Belin:2021nck} or the broader defect frameworks developed in~\cite{Benjamin:2025knd}. Such generalizations may lead to a nontrivial interplay between RCFT defects inherited from the seed theory and the universal permutation defects intrinsic to the orbifold structure.

We leave these directions for future investigation.

\acknowledgments
We are grateful to the Sepideh Mohammadi for useful comments and discussions.

%\newpage

\providecommand{\href}[2]{#2}\begingroup\raggedright\endgroup

\end{document}